\documentstyle[preprint,pra,aps]{revtex}
\tightenlines
\begin{document}
\title{\bf Calculation of positron binding to silver and gold atoms}
\author{V. A. Dzuba, V. V. Flambaum, and C. Harabati}
\address{School of Physics, The University of New South Wales, 
Sydney 2052, Australia}
\date{\today}
\maketitle

\begin{abstract}
Positron binding to silver and gold atoms was studied using a fully 
{\it ab initio} relativistic method, which combines the configuration
interaction method with many-body perturbation theory.
It was found that the silver atom forms a bound state with a
positron with binding energy 123 ($\pm$ 20\%) meV, while the gold
atom cannot bind a positron. Our calculations reveal the importance of
the relativistic effects for positron binding to heavy atoms.
The role of these effects was studied by varying the value
of the fine structure constant $\alpha$.
In the non-relativistic limit, $\alpha = 0$, both systems $e^+$Ag
and $e^+$Au are bound with binding energies of about 200 meV for $e^+$Ag
and 220 meV for $e^+$Au.
Relativistic corrections for a negative ion are essentially different
from that for a positron interacting with an atom. 
Therefore, the calculation of electron affinities cannot serve as a test of 
the method used for positron binding in the non-relativistic case. 
However, it is still a good test of the relativistic calculations.
Our calculated electron affinities for silver (1.327 eV) and
gold (2.307 eV) atoms are in very good agreement with corresponding
experimental values (1.303 eV and 2.309 eV respectively).
\end{abstract} 
\vspace{1cm}
\narrowtext
\section{Introduction}

Positron binding by neutral atoms has not been directly observed yet.
However, intensive theoretical study of the problem undertaken in the last
few years strongly suggests that many atoms can actually form bound states
with a positron (see, e.g. 
\cite{Dzuba95,Ryzhikh,RyMitVar98,Ryzhikh98,Mitroy99a,%
Mitroy99b,Stras98,Yuan98}). 
Most of the atoms studied so far were atoms with a relatively small value of 
the nuclear charge $Z$. It is important to extend the study to heavy atoms.
The main obstacle in this way is the rapid rise of computational 
difficulties with increasing number of electrons.
However, as we show in this paper, an inclusion of relativistic effects
is also important.
The role of these effects in positron binding to atoms has not been truly
appreciated. Indeed, one can say that due to strong Coulomb repulsion
a positron cannot penetrate to short distances from the nucleus and
remains non-relativistic.
However,
the positron binding is due to interaction with electrons which have large
relativistic corrections to their energies and wave functions.
The binding energy is the difference between the energies of a 
neutral atom and an atom bound with a positron. This difference is
usually small. On the other hand, relativistic contributions to the 
energies of both systems are large and there is no reason to expect
they are the same and cancel each other. Therefore, some relativistic
technique is needed to study positron binding by heavy atoms.

For both light and heavy atoms the main
 difficulty in calculations of positron interaction 
comes from the strong electron-positron Coulomb attraction.
This attraction leads to virtual positronium (Ps) formation \cite{Dzuba:93}. 
One can say that it gives rise
to a specific short-range attraction between the positron and the atom, in
addition to the usual polarizational potential which acts between a neutral
target and a charged projectile \cite{Dzuba95,Dzuba:93,Gribakin:94,Dzuba:96}.
 This
attraction cannot be treated accurately by perturbations and some all-order
technique is needed. In our earlier works 
\cite{Dzuba95,Dzuba:93,Gribakin:94,Dzuba:96}
we used the Ps wave function explicitly to approximate the virtual
Ps-formation contribution to the positron-atom interaction
and predicted $e^+$Mg, $e^+$Zn, $e^+$Cd and few other bound states.
The same physics
may also explain the success of the stochastic variational method in
positron-atom bound state calculations (see, e.g.  \cite{RyMitVar98} and Refs.
therein). In this approach the wave function is expanded in terms of
explicitly correlated Gaussian functions which include factors
$\exp (-\alpha r_{ij}^2)$ with inter-particle distances $r_{ij}$.
Using this method
Ryzhikh and Mitroy obtained positron bound states for a whole range of atoms
(Be, Mg, Zn, Cu, Ag, Li, Na, K, etc.). 
This method is well suited for few-particle systems. 
Its application to heavier systems is done by considering the
Hamiltonian of the valence electrons and the positron in the model potential
of the ionic core. However, for heavier atoms, e.g., Zn, the calculation
becomes extremely time consuming \cite{Mitroy99a}, and its convergence cannot
be ensured.

Another non-perturbative technique is the configuration interaction (CI) method
widely used in standard atomic calculations. This method was applied
to the positron-copper bound state in \cite{Mitroy99b}. In this work the
single-particle orbitals of the valence electron and positron are chosen as
Slater-type orbitals, and their interaction with the Cu$^+$ core is
approximated by the sum of the Hartree-Fock and model polarization potentials.
The calculation shows slow convergence with respect to the number of spherical
harmonics included in the CI expansion,
$L_{\max }=10$ being still not
sufficient to extrapolate the results reliably to $L_{\max }\rightarrow
\infty$.

In their more recent work the same authors applied the CI method to 
a number of systems consisting of an atom and a positron. These include
PsH, $e^+$Cu, $e^+$Li, $e^+$Be, $e^+$Cd and CuPs. In spite of some 
improvements to the method they still regard it as a ``tool with which to
perform preliminary investigations of positron binding'' \cite{Mitroy99}.

In our previous paper we developed a different version of the CI method
for the positron-atom problem \cite{Harabati}. 
The method is based on the relativistic Hartree-Fock method (RHF) and 
a combination of the CI method with  many body perturbation theory (MBPT). 
This method was firstly developed for pure electron systems \cite{Kozlov}
and its high effectiveness was demonstrated in a number of calculations 
\cite{Kozlov97,Johnson,Porsev}.
In the paper \cite{Harabati} it was successfully applied to the
positron binding by copper.
There are several important advances in the technique compared 
to the standard non-relativistic CI method which make it a very effective 
tool for the investigation of positron binding by heavy atoms.
\begin{enumerate}
\item The method is relativistic in the sense that the Dirac-Hartree-Fock
operator is used to construct an effective Hamiltonian for the problem
and to calculate electron and positron orbitals.
\item $B$-splines \cite{deBoor} in a cavity of finite radius $R$ were 
used to generate single-particle basis sets for an external electron and 
a positron.
The $B$-spline technique has the remarkable property of providing fast 
convergence with respect to the number of radial functions included
into the calculations \cite{Hansen,Sapirstein}. Convergence can be further
controlled by varying the cavity radius $R$ while the effect of the cavity 
on the energy of the system is taken into account analytically 
\cite{Harabati}. Convergence was clearly achieved for the 
$e^+$Cu system in Ref. \cite{Harabati} and for the
$e^+$Ag and $e^+$Au systems as presented below.
\item We use MBPT to include excitations from the core into the effective 
Hamiltonian. This corresponds to the inclusion of the correlations 
between core electrons and external particles (electron and positron)
and of the effect of screening of the electron-positron interaction
by core electrons. These effects are also often called the polarization of
the core by the external particles. We include them in a fully
{\it ab initio} manner up to the second order of the MBPT.
\end{enumerate}

In the present paper we apply this method to the problem of positron
binding by silver and gold atoms. Using a similar technique we also
calculate electron affinities for both these atoms. Calculations for
negative ions serve as a test of the technique used for positron-atom
binding. We also study the role of the relativistic effects in neutral
silver and gold, silver and gold negative ions and silver and gold 
interacting with a positron. This is done by varying the value of
the fine structure constant $\alpha$ towards its non-relativistic limit
$\alpha=0$.

\section{Theory}
\label{theory}

A detailed description of the method was given in Ref. \cite{Harabati}.
We briefly repeat it here emphasizing the role of the relativistic effects.
We use the relativistic Hartree-Fock method in the $V^{N-1}$ approximation
to obtain the single-particle basis sets of electron and positron orbitals
and to construct an effective Hamiltonian. 

The two-particle electron-positron wave function is given by the CI expansion,

\begin{equation}
	\Psi({\bf r}_e,{\bf r}_p)=
	\sum_{i,j}C_{ij} \psi^e_i({\bf r}_e) \psi^p_j({\bf r}_p),
\label{c_i}
\end{equation}
where $\psi^e_i$ and $\psi^p_j$ are the electron and positron orbitals
respectively. The expansion coefficients $C_{ij}$ are  determined by the
diagonalization of the matrix of the effective CI Hamiltonian acting in the
Hilbert space of the valence electron and the positron,
\begin{eqnarray}\label{HCI}
H_{\rm eff}^{\rm CI} &=& \hat h_e+\hat h_p + \hat h_{ep}, \nonumber \\
  \hat h_e &=&  c\bbox{\alpha p} + (\beta-1)mc^2
     - \frac{Ze^2}{r_e} + V_d^{N-1} - \hat V_{exch}^{N-1} + 
	\hat \Sigma_e, \nonumber \\
  \hat h_p  &=&  c\bbox{\alpha p} + (\beta-1)mc^2
    + \frac{Ze^2}{r_p} - V_d^{N-1} + \hat \Sigma_p, \\
\hat h_{ep} &=& - \frac{e^2}{|{\bf r}_e -{\bf r}_p|} + \hat \Sigma_{ep},
\nonumber 
\end{eqnarray}
where $\hat h_e$ and $\hat h_p$ are the effective single-particle 
Hamiltonians of
the electron and positron, and $\hat h_{ep}$ is the effective electron-positron
two-body interaction. Apart from the relativistic Dirac operator, $\hat h_e$
and $\hat h_p$ include the direct and exchange Hartree-Fock potentials of the
core electrons, $V_d^{N-1}$ and $V_{exch}^{N-1}$, respectively. The additional
$\hat \Sigma$ operators account for correlations involving core electrons.
$\Sigma_e$ and $\Sigma_p$ are single-particle operators which can be
considered as a self-energy part of the correlation interaction between
an external electron or positron and core electrons. 
These operators are often called ``correlation potentials'' due to the
analogy with the non-local exchange Hartree-Fock potential.
$\Sigma_{ep}$ 
represents the screening of the Coulomb interaction between external 
particles by core electrons
(see \cite{Harabati,Kozlov} for a detailed discussion). 

To study the role of the relativistic effects we use the form of the 
operators $h_e$ and $h_p$ in which the dependence on the fine structure 
constant $\alpha$ is explicitly shown.
Single-particle orbitals have the form
\begin{eqnarray}
\psi({\bf r})_{njlm} = \frac{1}{r}
 \left( \begin{array}{c}
f_n(r) \Omega({\bf r}/r)_{jlm} \\ i 
\alpha g_n(r) \tilde{\Omega}({\bf r}/r)_{jlm}
\end{array}
\right).
\label{psi}
\end{eqnarray}
Then the RHF equations 
\[
(h_i-\epsilon_n)\psi^i_n = 0, \ \ (i=e,p)
\]
take the following form
\begin{eqnarray}\label{RHF}
f'_n(r)+\frac{\kappa_n}{r}f_n(r)-[2+\alpha^2(\epsilon_n - \hat V)]g_n(r) = 0\\
g'_n(r)-\frac{\kappa_n}{r}g_n(r)+(\epsilon_n - \hat V)f_n(r) = 0 \nonumber ,
\end{eqnarray}
where $\kappa = (-1)^{l+j+1/2}(j+1/2)$ and $V$ is the effective potential
which is the sum of the Hartree-Fock potential and correlation potential 
$\Sigma$:
\begin{eqnarray}
\hat V & = & -\frac{Ze^2}{r_e} +V_d^{N-1} - \hat V_{exch}^{N-1}+\hat \Sigma_e,
\ \ \mbox{- for an electron}, \nonumber \\
\hat V & = &  \frac{Ze^2}{r_p} -V_d^{N-1} +\hat \Sigma_p,
\ \ \mbox{- for a positron}.
\label{hfpot}
\end{eqnarray}
The non-relativistic limit can be achieved by reducing the value of
$\alpha$ in (\ref{RHF}) to $\alpha = 0$.

The relativistic energy shift in atoms with
one external electron can also be estimated by the following
equation \cite{Webb}
\begin{eqnarray}
	\Delta_n = \frac{E_n}{\nu}(Z\alpha)^2 \left[\frac{1}{j+1/2} -
	C(Z,j,l)\right],
\label{rel5}
\end{eqnarray}
where $E_n$ is the energy of an external electron, $\nu$ is the effective
principal quantum number ($E_n = -0.5/\nu^2$ a.u.). The coefficient
$C(Z,j,l)$ accounts for many-body effects.
Note that formula (\ref{rel5}) is based on the specific expression
for the electron density in the vicinity of the nucleus and therefore
is not applicable for a positron. 

\section{Silver and gold negative ions}\label{se:Ag-}

We calculated electron affinities of silver and gold atoms mostly to 
test the technique used for positron-atom binding. The calculation of a
negative ion Ag$^-$ or Au$^-$ is a two-particle problem technically
very similar to positron-atom binding. 
The effective Hamiltonian of the problem has a form similar to (\ref{HCI})
\begin{eqnarray}\label{HCIe}
	H_{\rm eff}^{\rm CI} &=& \hat h_e(r_1)+\hat h_e(r_2) + \hat h_{ee}, 
	\nonumber \\
\hat h_{ee} &=& \frac{e^2}{|{\bf r}_e -{\bf r}_p|} + \hat \Sigma_{ee},
\nonumber 
\end{eqnarray}
where $ \hat \Sigma_{ee}$ represents the screening of the Coulomb interaction
between external electrons by core electrons (see Refs. \cite{Kozlov,Harabati}
for detailed discussion).
Electron affinity is defined
when an electron can form a bound state with an atom. In this case the 
difference between the energy of a neutral atom and the energy of a
negative ion is called the electron affinity to this atom.
Energies of Ag, Ag$^-$, Au, Au$^-$ obtained in different approximations
and corresponding electron affinities are presented in Table \ref{Cum}
together with experimental data. The energies are given with respect 
to the cores (Ag$^+$ and Au$^+$).
Like in the case of Cu$^-$ \cite{Harabati} the accuracy of the Hartree-Fock
approximation is very poor.  The 
binding energies of the $5s$ electron in neutral Ag and the $6s$ electron in 
neutral Au are underestimated by
about 21\% and 23\% respectively, while the negative ions are unbound.
Inclusion of either core-valence correlations ($\Sigma$) or valence-valence
correlations (CI) does produce binding but the accuracy is still poor.
Only when both these effects are included the accuracy for the electron 
affinities improves significantly becoming 20\% for Ag$^-$ and 11\% 
for Au$^-$. Further improvement can be achieved  by introducing numerical 
factors before $\hat \Sigma_e$ to fit the lowest $s,p$ and $d$ energy levels 
of the neutral atoms. These factors simulate the effect of higher-order
correlations.
Their values are $f_s = 0.88$,
$f_p = 0.97$, $f_d = 1.08$ for the Ag atom and $f_s = 0.81$,
$f_p = 1$, $f_d = 1.04$  for the Au atom in the $s$, $p$ and $d$ channels, 
respectively. As is evident from Table \ref{Cum}, the fitting of the
energies of neutral atoms also significantly improves electron affinities.
It is natural to assume that the same procedure should work equally well for 
the positron-atom problem. 

Results of other calculations of the electron affinities of silver and gold 
are presented in Table \ref{others} together with the experimental values.

\section{Positron binding to silver and gold and the role of relativistic 
effects}\label{se:Ag+}

As for the case of copper \cite{Harabati} we have performed calculations
for two different cavity radii $R=30a_0$ and $R=15a_0$. For a smaller radius
convergence with respect to the number of single-particle basis states
is fast. However, the effect of the cavity on the 
converged energy is large. For a larger cavity radius, convergence is slower
and the effect of the cavity on the energy is small. 
When the energy shift caused
by the finite cavity radius is taken into account both calculations come to
the same value of the positron binding energy. Table \ref{silver} illustrates
the convergence of the calculated energies of $e^+$Ag and $e^+$Au with respect
to the maximum value of the angular momentum of single-particle orbitals.
Energies presented in the table are two-particle
energies (in a.u.) with respect to the energies of Ag$^+$ and Au$^+$.
The number of radial orbitals $n$ in each partial wave is fixed at $n=16$.
Fig. 1 shows the convergence of the calculated energy with respect to $n$ 
when maximum momentum of the single-particle orbitals was fixed at $L=10$.
The cavity radius in both cases was $R=30a_0$. Table \ref{silver} and Fig. 1 
show that even for a larger cavity radius, convergence was clearly achieved. 
Table \ref{silver} also shows the convergence in different approximations,
namely with and without core-valence correlations ($\Sigma$). One can see
that while inclusion of $\Sigma$ does shift the energy, the convergence is 
not affected.

Table \ref{final} shows how positron binding by silver and gold is formed 
in different approximations. This table is very similar to Table \ref{Cum}
for the negative ions except there is no RHF approximation for the
positron binding. Indeed, the RHF approximation for the negative ions means 
a single-configuration approximation: $5s^2$ for Ag$^-$ and $6s^2$ for Au$^-$.
These configurations strongly dominate in the two-electron wave function
of the negative ions even when a large number of configurations are mixed
to ensure convergence. In contrast, no single configuration strongly
dominates in the positron binding problem. Therefore we present our results 
in Table \ref{final} starting from the standard CI approximation.
In this approximation positron is bound to both silver and gold atoms.
However, the inclusion of core-valence correlations through the introduction
of the $\Sigma_e$, $\Sigma_p$ and $\Sigma_{ep}$ operators shifts the energies
significantly. In the case of gold, the $e^+$Au system becomes unbound when
all core-valence correlations are included.

As was discussed in our previous paper \cite{Harabati} the dominating factor
affecting the accuracy of the calculations is higher-order correlations
which mostly manifest themself via the value of the $\Sigma$ operator.
An introduction of the fitting parameters  as described
in the previous section can be considered as a way to simulate the effect
of higher-order correlations. Also, the energy shift caused by the
fitting can be considered as an estimation of the uncertainty of
the calculations. This shift is 0.00240 a.u. in the case of silver and 
0.00023 a.u. in the case of gold (see Table \ref{final}). Note that
these values are considerably smaller than energy shifts for the
silver and gold negative ions (0.00854 a.u. and 0.00921 a.u. 
respectively, see Table \ref{Cum}). This is because of the cancellation of
the effects of the variation of $\Sigma_e$ and $\Sigma_p$. In particular,
for gold it is accidentally very small.
One can see that even if the value of 0.00240 a.u. is adopted as an 
upper limit of the uncertainty of the calculations, the
$e^+$Ag system remains bound while the $e^+$Au system remains unbound.
However, the actual accuracy might be even higher. We saw that the
fitting procedure significantly improves the accuracy of the calculations 
for the silver and gold negative ions. It is natural to assume that the
same procedure works equally well for the positron binding problem.
The final result for the energy of positron binding by the silver atom as
presented in Table \ref{final} is 0.00434 a.u. This result does not
include the effect of the finite cavity size. When this effect is taken
into account, by means of the procedure described in Ref. \cite{Harabati},
the binding energy becomes 0.00452 a.u. or 123 meV. If we adopt the value
of 0.00240 a.u as an estimation of the uncertainty of the result, then the 
accuracy we can claim is about 20\%.

The calculation of the positron binding by copper \cite{Harabati}, silver and 
gold reveal an interesting trend. All three atoms have very similar
electron structure. However the positron binding energy for silver (123 meV)
is considerably smaller than that for copper (170 meV \cite{Harabati})
while gold atoms cannot bind positrons at all. We believe that this trend 
is caused by relativistic effects. An argument that the positron is always 
non-relativistic does not look very convincing because electrons also
contribute to the binding energy. Relativistic effects are large for
heavy atoms and electron contributions to the positron binding energy
could be very different in the relativistic and non-relativistic limits.
Indeed, we demonstrated in Ref. \cite{Webb} that the relativistic
energy shift considerably changes the values of the transition frequencies
in Hg$^+$ ion and sometimes even changes the order of the energy levels.
If we use formula (\ref{rel5})
with the contribution of the many-body effects $C=0.6$,
as suggested in Ref. \cite{Webb}, to estimate the relativistic 
energy shift for neutral Au then the result is -0.037 a.u. This is 
about an order of magnitude larger than the energy difference between
Au and $e^+$Au.
If the relativistic energy
shift in $e^+$Au is  different from that in Au then
the positron binding energy may be strongly affected.

To study the role of the relativistic effects in positron binding in more
detail we performed the calculations for Ag, Ag$^-$, $e^+$Ag,
Au, Au$^-$ and $e^+$Au in the relativistic and non-relativistic
limits. The latter corresponds to the zero value of the fine structure 
constant $\alpha$ (see Section \ref{theory}).
The results are presented in Table \ref{rel}. One can see that the actual
relativistic energy shift for neutral Au is even bigger than is
suggested by formula (\ref{rel5}) with $C=0.6$. The shift is 0.0805 a.u.
which corresponds to $C=0.08$. Formula (\ref{rel5}) with $C=0.08$ also
reproduces the relativistic energy shift for neutral Ag. The relativistic
energy shift for an atom with a positron is of the same order of
magnitude but a little different in value. This difference turned out
to be enough to affect the positron binding energy significantly.
In particular, the $e^+$Au system which is unbound in relativistic
calculations becomes bound in the non-relativistic limit with binding 
energy 0.0080 a.u or 218 meV. In the case of silver, the positron binding
energy is considerably higher in the non-relativistic limit.
It is 0.0073 a.u. or 199 meV. It is interesting to compare this value
with the value of 150 meV obtained by Mitroy and Ryzhikh using
the non-relativistic stochastic variational method \cite{Ryzhikh98}.
Since the convergence was achieved in both calculations the remaining
difference should probably be attributed to the different treatment
of the core-valence correlations. We use many-body perturbation theory
for an accurate calculation of the $\Sigma$ operator which accounts for
these correlations. Mitroy and Ryzhikh use an approximate semi-empirical
expression for the $\Sigma$ operator which is based on its long-range
asymptotic behavior.

Note that the relativistic energy shift for negative ions is also large.
However electron affinities are less affected. This is because electron
affinities are many times larger than positron binding energies and
therefore less sensitive to the energy shift. Apart from that there is
a strong cancellation between relativistic energy shifts in the negative 
ion and neutral atom.
This means in particular that the calculation
of the electron affinities cannot serve as a test of a non-relativistic
method chosen for the positron binding problem. However, it is still a
good test of the relativistic calculations.
Note also that our calculated relativistic energy shifts for neutral and
negative silver and gold are in very good 
agreement with calculations performed by Schwerdtfeger and Bowmaker by
means of relativistic and non-relativistic versions of the quadratic
configuration interaction method (see Table \ref{relc} and Ref. 
\cite{Bowmaker}).

The authors are grateful to G. F. Gribakin for many useful discussions.



\begin{table}
\caption{Ground state energies (in a.u.) of silver, gold and their negative 
ions calculated in different approximations}
\label{Cum}
\begin{tabular}{lddd}
 & Neutral atom & Negative ion & Electron affinity\tablenotemark[1] \\
\hline
 &\multicolumn{3}{c}{Silver}  \\
RHF\tablenotemark[2]            & -0.22952 & -0.20156 & -0.02795 \\
RHF + $\Sigma$\tablenotemark[3] & -0.27990 & -0.30231 &  0.02241 \\
CI\tablenotemark[4]             & -0.22952 & -0.25675 &  0.02722 \\
CI +$\Sigma_e$\tablenotemark[5] & -0.28564 & -0.33560 &  0.04996 \\
CI +$\Sigma_e + \Sigma_{ee}$\tablenotemark[6] 
                                & -0.28564 & -0.34298 &  0.05734 \\
CI + $f \Sigma_e + \Sigma_{ee}$\tablenotemark[7]        
                                & -0.27841 & -0.32721 &  0.04880 \\
Experiment\tablenotemark[8]     & -0.27841 & -0.32626 &  0.04784 \\
 &\multicolumn{3}{c}{Gold} \\
RHF\tablenotemark[2] & -0.27461 & -0.26169 & -0.01292 \\
RHF + $\Sigma$\tablenotemark[3] & -0.34900 & -0.41046 &  0.06146 \\
CI\tablenotemark[4]             & -0.27461 & -0.31369 &  0.03908 \\
CI +$\Sigma_e$\tablenotemark[5] & -0.35536 & -0.43913 &  0.08376 \\
CI +$\Sigma_e + \Sigma_{ee}$\tablenotemark[6]   
                                & -0.35536 & -0.44943 &  0.09407 \\
CI + $f \Sigma_e + \Sigma_{ee}$\tablenotemark[7]        
                                & -0.33903 & -0.42389 &  0.08486 \\
Experiment\tablenotemark[8]     & -0.33903 & -0.42386 &  0.08483 \\
\end{tabular}
\tablenotetext[1]{Negative affinity means no binding.}
\tablenotetext[2]{Relativistic Hartree-Fock; a single-configuration
approximation, no core-valence correlations are included.}
\tablenotetext[3]{Single-configuration approximation, core-valence 
correlations are included by means of MBPT.}
\tablenotetext[4]{Standard CI method.}
\tablenotetext[5]{Self-energy part of core-valence correlations are
included by adding the $\Sigma_e$ operator to the CI Hamiltonian.} 
\tablenotetext[6]{CI+MBPT method, self-energy and screening correlations
are included by $\Sigma$ operators while valence-valence
correlations are included by configuration interaction.}
\tablenotetext[7]{$\Sigma_e$ in different waves are taken with factors 
to fit energies of a neutral atom.}
\tablenotetext[8]{References \cite{Moore,Miller}.}
\end{table}
\begin{table}
\caption{Electron affinities of Ag and Au (eV). Comparison with other 
calculations and experiment.}
\label{others}
\begin{tabular}{ddll}
Ag & Au & Ref. & Method\\
\hline
\multicolumn{3}{c}{Theory} & \\
1.008 &1.103 & \cite{Bowmaker} & Non-relativistic quadratic configuration
interaction method\\
1.199 &2.073 & \cite{Bowmaker} & Relativistic quadratic configuration 
interaction method\\
1.254 &2.229 & \cite{Neogrady} & Relativistic coupled cluster method  \\
1.022 & & \cite{Ryzhikh98} & Non-relativistic stochastic variational method\\
& 2.28 & \cite{Eliav} &  Fock-space relativistic coupled-cluster method\\
& 2.26 & \cite{Kaldor} & Fock-space coupled-cluster method with Douglas-Kroll\\
 & & & transformation (relativistic) \\
1.327 & 2.307 & & Present work \\
\multicolumn{3}{c}{Experiment} & \\
1.303 & 2.309 & \cite{Hotop}& \\
\end{tabular}
\end{table}
\begin{table}
\caption{Convergence of the calculation of the energies of $e^+$Ag
and $e^+$Au with respect to the number of included partial waves (a.u.)}
\label{silver}
\begin{tabular}{ldccc}
 & $L_{\rm max}$ & CI\tablenotemark[1] & CI +$\Sigma$\tablenotemark[2] & 
CI + $f \Sigma$\tablenotemark[3] \\
\hline
$e^+$Ag & 0  & $-$0.2232729 & $-$0.2800223 & $-$0.2729038 \\
        & 1  & $-$0.2271709 & $-$0.2838360 & $-$0.2749591 \\
        & 2  & $-$0.2309207 & $-$0.2868375 & $-$0.2765124 \\
        & 3  & $-$0.2350823 & $-$0.2895691 & $-$0.2780571 \\
        & 4  & $-$0.2388315 & $-$0.2916800 & $-$0.2793784 \\
        & 5  & $-$0.2419251 & $-$0.2932381 & $-$0.2804487 \\
        & 6  & $-$0.2443218 & $-$0.2943470 & $-$0.2812678 \\
        & 7  & $-$0.2460745 & $-$0.2951085 & $-$0.2818603 \\
        & 8  & $-$0.2472812 & $-$0.2956100 & $-$0.2822647 \\
        & 9  & $-$0.2480477 & $-$0.2959189 & $-$0.2825199 \\
        & 10 & $-$0.2484749 & $-$0.2960829 & $-$0.2826596 \\
        & 11 & $-$0.2486698 & $-$0.2961444 & $-$0.2827143 \\
        & 12 & $-$0.2487554 & $-$0.2961682 & $-$0.2827367 \\
        & 13 & $-$0.2487928 & $-$0.2961778 & $-$0.2827459 \\
        & 14 & $-$0.2488090 & $-$0.2961817 & $-$0.2827498 \\
\hline
$e^+$Au & 0  & $-$0.2684049 & $-$0.3500447 & $-$0.3330163 \\
        & 1  & $-$0.2706582 & $-$0.3526602 & $-$0.3339500 \\
        & 2  & $-$0.2719813 & $-$0.3539745 & $-$0.3344564 \\
        & 3  & $-$0.2732705 & $-$0.3550481 & $-$0.3348765 \\
        & 4  & $-$0.2743905 & $-$0.3558030 & $-$0.3351787 \\
        & 5  & $-$0.2753222 & $-$0.3563289 & $-$0.3353973 \\
        & 6  & $-$0.2760539 & $-$0.3566883 & $-$0.3355525 \\
        & 7  & $-$0.2765943 & $-$0.3569283 & $-$0.3356590 \\
        & 8  & $-$0.2769686 & $-$0.3570837 & $-$0.3357294 \\
        & 9  & $-$0.2772074 & $-$0.3571791 & $-$0.3353733 \\
        & 10 & $-$0.2773390 & $-$0.3572293 & $-$0.3357972 \\
        & 11 & $-$0.2773925 & $-$0.3572449 & $-$0.3358049 \\
        & 12 & $-$0.2774146 & $-$0.3572505 & $-$0.3358078 \\
        & 13 & $-$0.2774239 & $-$0.3572527 & $-$0.3358091 \\
        & 14 & $-$0.2774278 & $-$0.3572536 & $-$0.3358095 \\
\end{tabular}
\tablenotetext[1]{Standard CI method.}
\tablenotetext[2]{CI+MBPT method, both core-valence and valence-valence
correlations are included.}
\tablenotetext[3]{$\Sigma$ is taken with fitting parameters as explained 
in the text.}
\end{table}

\begin{table}
\caption{Positron binding by silver and gold
 calculated in different approximations (all energies are in a.u.)}
\label{final}
\begin{tabular}{lddd}
 & Neutral atom & Atom with $e^+$ & $\Delta$\tablenotemark[1] \\
\hline
 &\multicolumn{3}{c}{Silver} \\
CI                        & -0.22952 & -0.24881 &  0.01929 \\
CI +$\Sigma_e + \Sigma_p$ & -0.28564 & -0.29618 &  0.01054 \\
CI +$\Sigma_e + \Sigma_p + \Sigma_{ep}$
                          & -0.28564 & -0.28843 &  0.00279 \\
CI + $f \Sigma_e + f \Sigma_p + \Sigma_{ep}$
                          & -0.27841 & -0.28275 &  0.00434 \\
 &\multicolumn{3}{c}{Gold} \\
CI                        & -0.27461 & -0.27743 &  0.00282 \\
CI +$\Sigma_e + \Sigma_p$ & -0.35536 & -0.35725 &  0.00189 \\
CI +$\Sigma_e + \Sigma_p + \Sigma_{ep}$
                          & -0.35536 & -0.35191 &  -0.00345 \\
CI + $f \Sigma_e + f \Sigma_p + \Sigma_{ep}$
                          & -0.33903 & -0.33581 &  -0.00322 \\
\end{tabular}
\tablenotetext[1]{Positron binding energy. Negative energy means no binding.}
\end{table}
\begin{table}
\caption{Energies (in a.u.) of Ag, Ag$^-$, $e^+$Ag, Au, Au$^-$ 
and $e^+$Au with respect to the energy of the core in relativistic 
and non-relativistic cases}
\label{rel}
\begin{tabular}{cddddd}
 & Neutral & Negative & Atom with  & Electron & Positron binding \\
 &  atom   &   ion    & a positron & affinity &  energy\tablenotemark[1] \\
\hline
  \multicolumn{6}{c}{Silver}\\
Non-relativistic   & -0.2558 & -0.2974 & -0.2640 & 0.0416 & 0.0073 \\
Relativistic       & -0.2784 & -0.3272 & -0.2827 & 0.0488 & 0.0043 \\
$\Delta$ &            0.0226 &  0.0298 &  0.0187 &-0.0072 & 0.0030 \\
  \multicolumn{6}{c}{Gold}\\
Non-relativistic   & -0.2537 & -0.3040 & -0.2665 & 0.0503 & 0.0080 \\
Relativistic       & -0.3390 & -0.4239 & -0.3358 & 0.0849 &-0.0032 \\
$\Delta$ &            0.0853 &  0.1199 &  0.0693 &-0.0346 & 0.0112 \\
\end{tabular}
\tablenotetext[1]{Positive energy means bound state}
\end{table}
\begin{table}
\caption{Comparison of the relativistic energy shift with other calculations
(energies are in a.u.)}
\label{relc}
\begin{tabular}{ldd}
Atom/Ion & Present work & Schwerdtfeger and Bowmaker \tablenotemark[1]\\
\hline
Ag     & 0.0226 & 0.0200 \\
Ag$^-$ & 0.0072 & 0.0070 \\
Au     & 0.0853 & 0.0714 \\
Au$^-$ & 0.0346 & 0.0357 \\
\end{tabular}
\tablenotetext[1]{Quadratic configuration interaction method, Ref. 
\cite{Bowmaker}}
\end{table}
\widetext
\newpage
\input psfig
\psfull
\begin{figure}[b]
\psfig{file=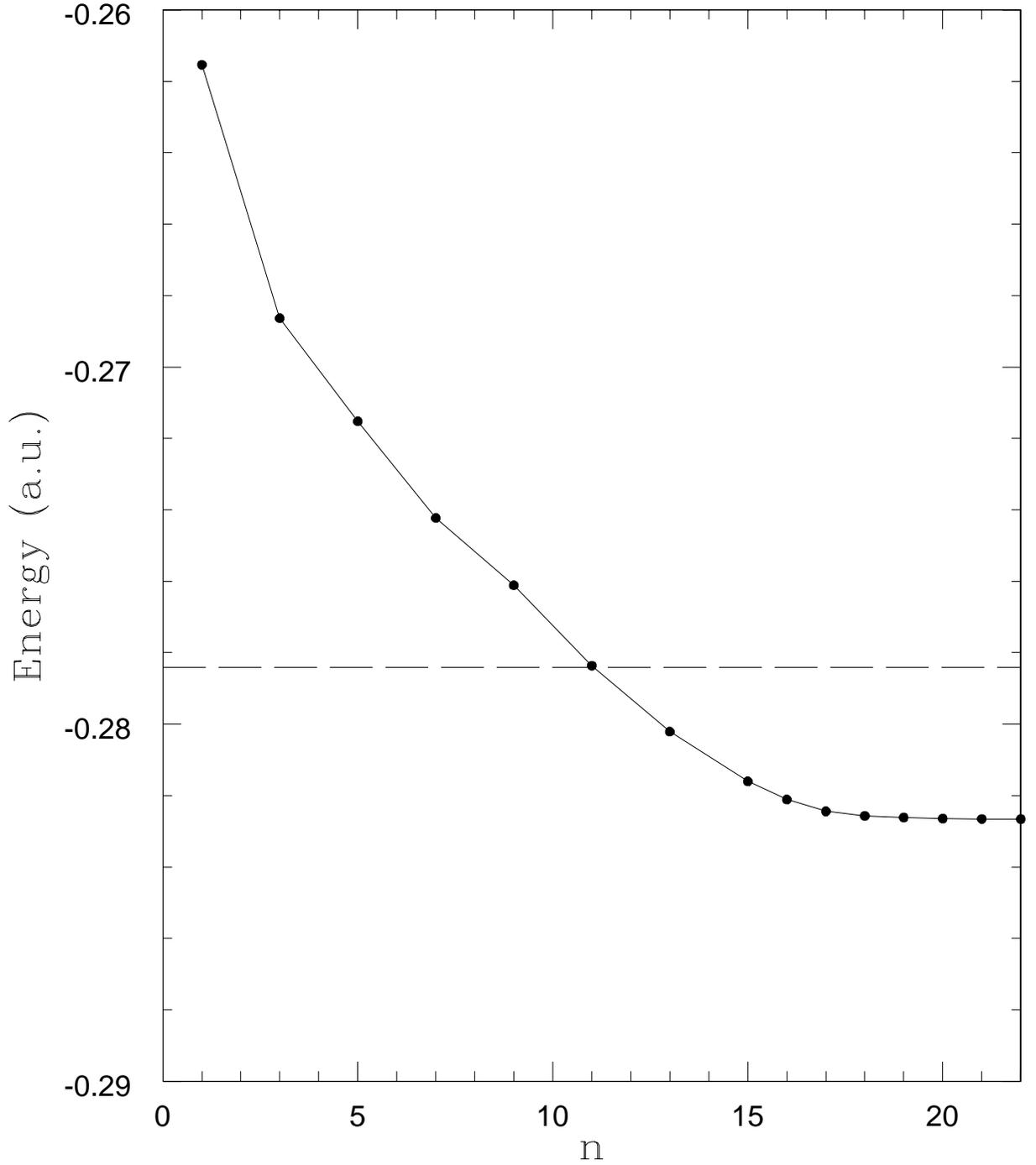, clip=}
\caption{Energy of $e^+$Ag as a function of the number of radial electron
and positron basis functions in each partial
wave ($L_{\rm max}= 10$) in the cavity with $R=30a_0$. Dashed line represents
the energy of neutral silver.}
\label{Agplot}
\end{figure}
\begin{figure}[b]
\psfig{file=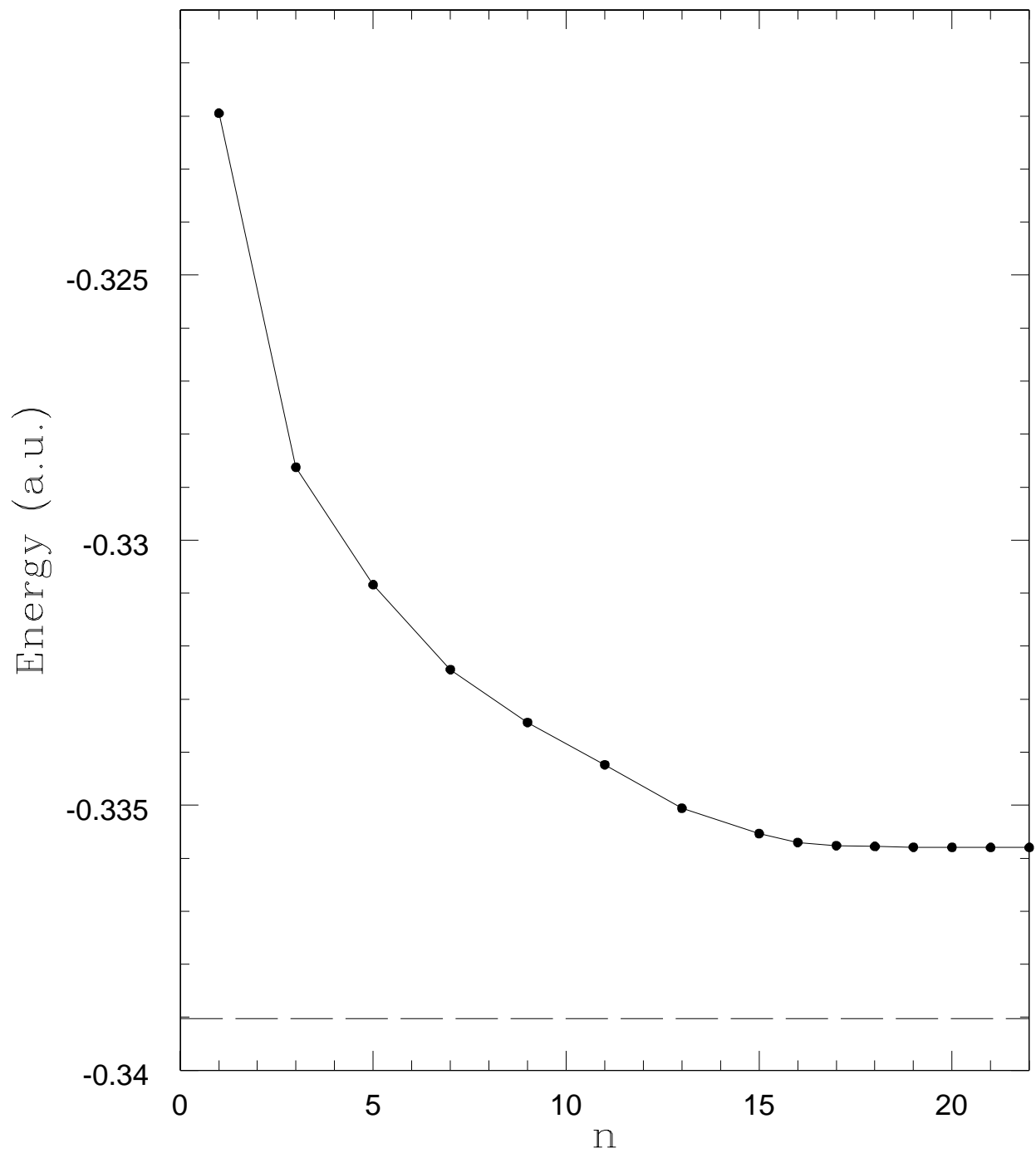, clip=}
\caption{Same as Fig. 1 but for $e^+$Au.}
\label{Auplot}
\end{figure}

\end{document}